\documentclass[twoside,twocolumn,english,superscriptaddress]{revtex4-1}

\usepackage{color}
\usepackage{float}
\usepackage{amsmath}
\usepackage{graphicx}

\makeatletter

\usepackage[caption=false]{subfig}

\begin{document}
\begin{center}
\date{9-3-2020}
\end{center}

\title{Models for the yielding behaviour of amorphous solids}

\author{Srikanth Sastry}
\email{sastry@jncasr.ac.in}
\affiliation{Jawaharlal Nehru Centre for Advanced Scientific Research, Jakkur Campus, Bengaluru 560064, India.}

\captionsetup[subfigure]{labelformat=empty}

\newpage
\newpage

\begin{abstract}
Understanding the mechanical response and failure of solids is of obvious importance in their use as structural materials. The nature of plastic deformation leading to yielding of amorphous solids has been vigorously pursued in recent years. Investigations employing both unidirectional and cyclic deformation protocols reveal a strong dependence of yielding behaviour on the degree of annealing. Below a threshold degree of annealing, the nature of yielding changes qualitatively, to progressively more discontinuous yielding.  Theoretical investigations of yielding in amorphous solids have almost exclusively focused on yielding under unidirectional deformation, but cyclic deformation reveals several interesting features that remain largely un-investigated. Focusing on athermal cyclic deformation, I investigate a family of models based on an energy landscape description. These models reproduce key interesting features observed in simulations, and provide an interpretation for the intriguing presence of a threshold energy.
\end{abstract}
\maketitle

Amorphous solids of a wide variety are of scientific and technological importance. Apart from {\it hard glasses} such as oxide glasses, metallic glasses, {\it etc.}, several {\it soft materials}, such as colloidal assemblies, gels, emulsions and pastes  also belong to the broad spectrum of amorphous solids \cite{Bonn2017c,Nicolas2018}.
The manner in which such solids respond to external stresses is equally diverse, understanding of which is of interest for many reasons --  whereas the development of plastic deformation leading to mechanical failure may be a phenomenon to predict and prevent in the case of structural materials, controlling the elastoplastic {\it flow} properties is of interest in the case of soft materials, such as gels and pastes \cite{Bonn2017c}. A general description of the mechanical response of amorphous solids needs to take note of the microscopic structural disorder and account for the apparent diversity of responses. 

The nature of elementary processes of plasticity, interactions between them, the nature of the yielding transition and the dependence of the mechanical response on preparation history, have been actively investigated over the years \cite{Falk2011,Dasgupta2012,JinWyart2016,regev2015reversibility,Itamar2016yielding,Jin,parisi2017shear,Urbani2017b,leishangthem2017yielding,Ozawa2018a,Bhaumik2019,Radhakrishnan2016b,Popovic2018a,Fielding2020}. In particular, computer simulations of {\it athermal quasi-static shear} (AQS) deformations \cite{shi2007,leishangthem2017yielding,Ozawa2018a,Bhaumik2019,Yeh2019,VishwasPRE2020,PRIEZJEV2018}, employing both uniform and cyclic shear protocols, reveal that the character of yielding depends strongly on the initial degree of annealing of the solid. Under uniform shear, the evolution of stress with strain is gradual for poorly annealing glasses, whereas well annealed glasses display both a stress overshoot and a discontinuous jump down to the flow stress, with the strength of discontinuity growing with annealing, below a threshold. Such variation of yielding behaviour with annealing has been suggested as a framework to understand the observed diversity among amorphous solids \cite{Ozawa2018a}. 

\begin{figure}[htp]
\centering{}
\includegraphics[scale=0.22]{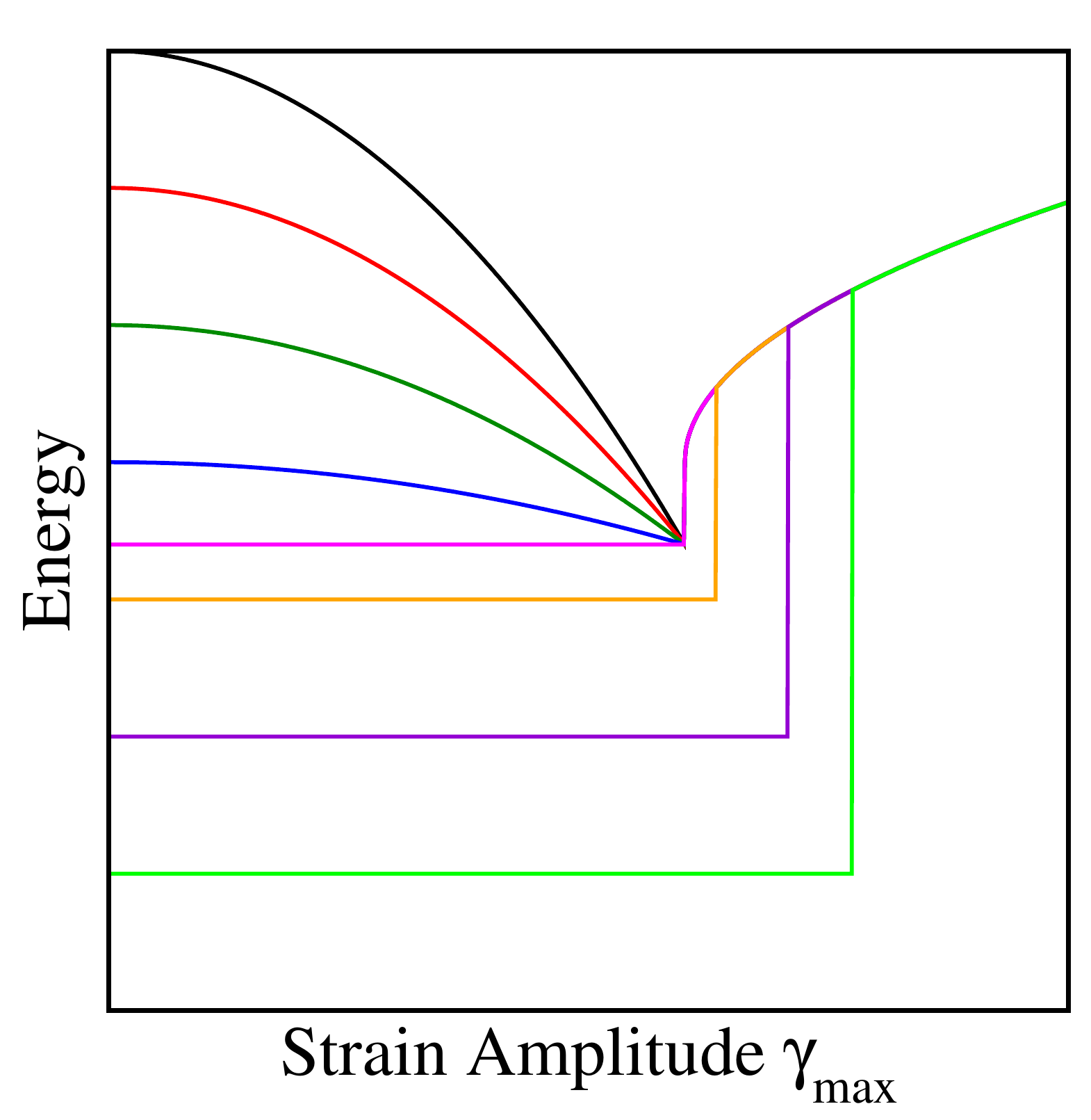}
\includegraphics[scale=0.22]{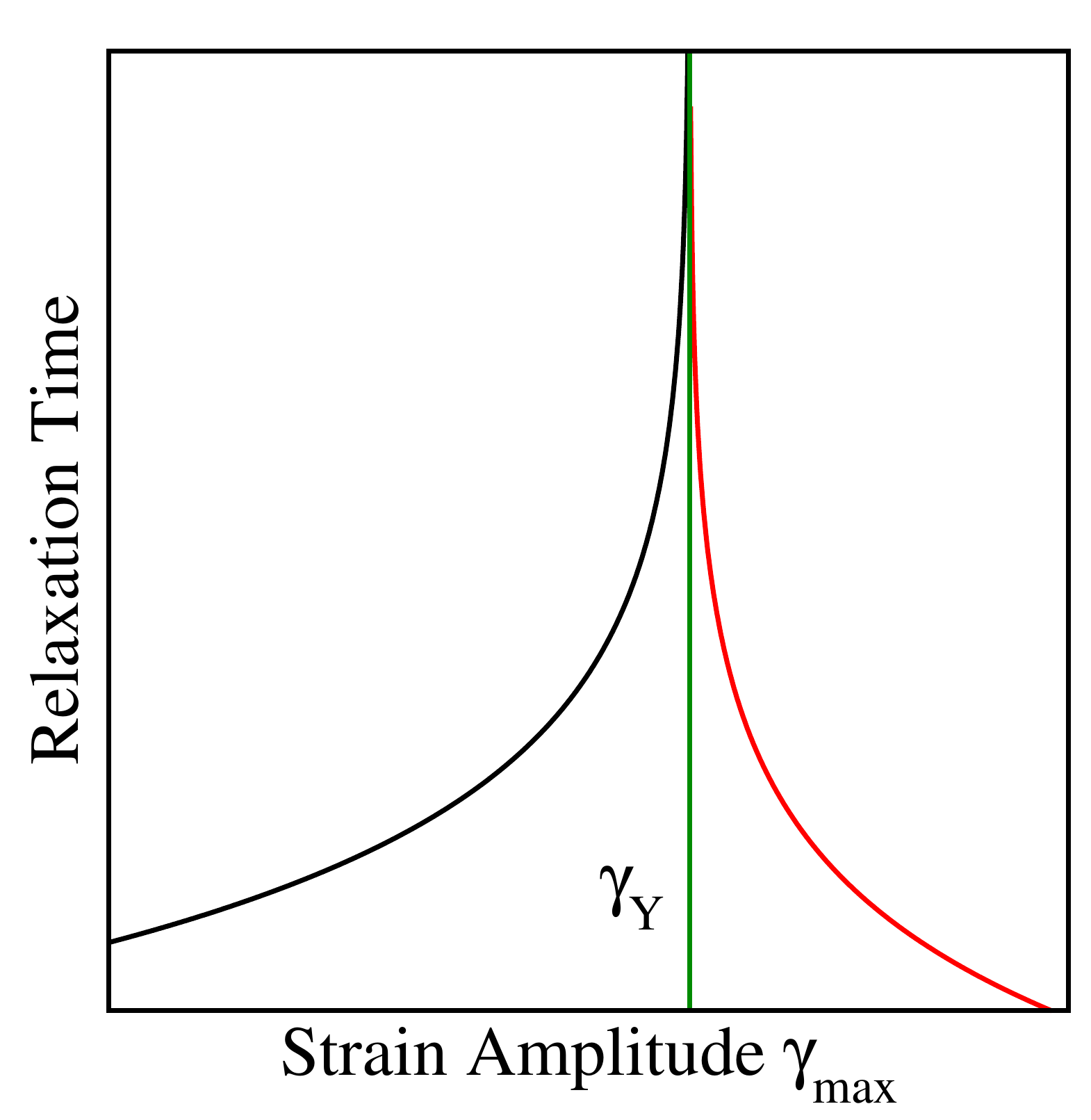} 
\includegraphics[scale=0.5]{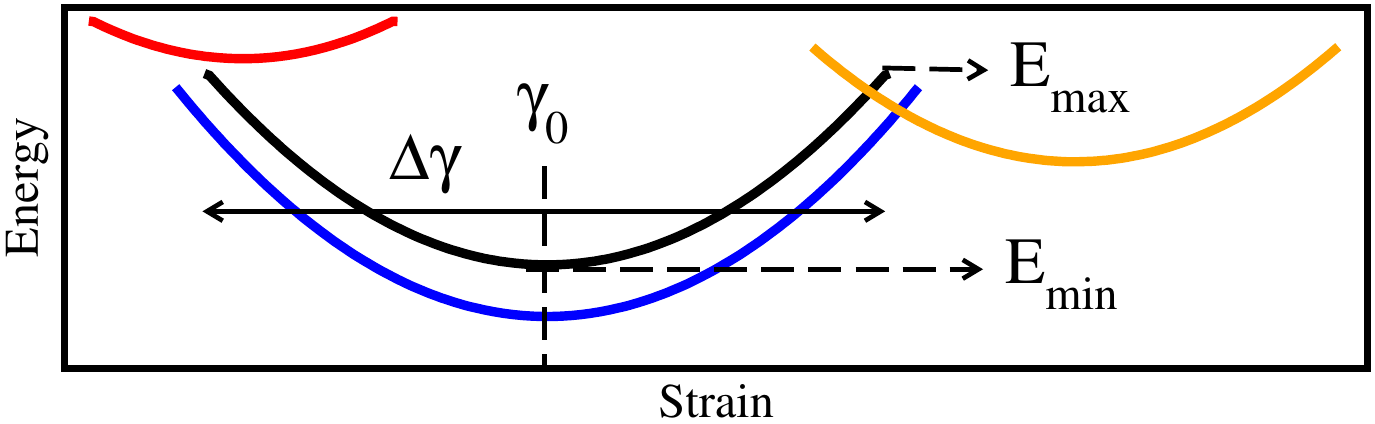} 
\caption{(Top) Schematic representation of yielding behaviour under cyclic shear. (a) (Left) Dependence of steady state energy on strain amplitude $\gamma_{max}$ for different initial energies. (b) (Right) Time required to reach the steady state {\it vs.} strain amplitude $\gamma_{max}$. (Bottom) A schematic description of the model. The minimum ($E_{min}$) and maximum ($E_{max}$) energy, strain at which energy is minimum ($\gamma_0$) and the stability range ($\Delta \gamma$) are indicated for one mesostate (black), from which transitions are possible to two other mesostates shown (blue, orange) but not to the third (red). } 
\label{fig:1}
\end{figure}

In the case of cyclic shear, poorly annealed glasses, above a threshold degree of annealing (or energy), anneal towards a threshold energy with increasing amplitude of strain, before yielding \cite{leishangthem2017yielding,Bhaumik2019,Yeh2019}, accompanied by shear banding \cite{parmar2019strain} (as also for uniform shear). Initially well annealed samples with energies below the threshold do not show any change in properties untill they yield discontinuously. Above yielding, the properties do not depend on the initial sample history. Interestingly, the number of cycles required to reach the steady state increases strongly as the strain amplitude approaches, and possibly diverges at, the yielding amplitude\cite{Fiocco2013,regev2015reversibility,leishangthem2017yielding}. These features are schematically summarised in Fig. \ref{fig:1}. 

Based on the observations that plastic deformation involves spatially localised arrangements (termed {\it shear transformation zones}), and that they lead to the generation of long range stress fields, a variety of {\it elasto-plastic} models \cite{Nicolas2018} have been developed and studied, aiming to capture macroscopic deformation response, through a combination of a coarse-grained description of local plasticity, and a continuum description at larger length scales. Attempts have been made to incorporate the role of annealing in such models, for uniform shear \cite{Talamali2012,Ozawa2018a,Popovic2018a,Fielding2020}, but response to cyclic shear remains largely unaddressed. In particular, features exhibited by athermal cyclic shear, including mechanically induced aging or annealing, the presence of a threshold energy across which the character of yielding changes, the apparent divergence  of time scales have not been demonstrated to arise in such models. One may well demand that a satisfactory elasto-plastic model be able to capture such phenomena. A desirable model should capture both athermal and thermal effects, and have the potential to incorporate parametrically properties that may be specific to a variety of amorphous solids. The ability to capture memory effects seen in cyclically sheared glasses would be another desirable feature \cite{keimMemoryReview2019,Mungan2019}. Such expectations point to an energy based model \cite{sollich1997rheology,jagla2007strain,Merabia_2016,jagla2017}) (such as the soft glassy rheology or SGR model \cite{sollich1997rheology}), rather than models defined in terms of threshold stresses, as with typical elasto-plastic models. 

The models studied here are defined in terms of the distribution of states or energy minima that a mesoscale block of an amorphous solid can be in, and rules for transitions between them. Inclusion of an extended array for such blocks and interactions between them along previously developed lines  \cite{Nicolas2018} is an obvious next step but is not pursued here. Each state (termed a {\it mesostate} and investigated in detail in \cite{Mungan2019}) is characterised by (i) an energy $E_0$ (taken to be negative always) at a strain value $\gamma_0$ at which the energy is minimum, (ii) a stability range in strain over which it is stable, and (iii) a form for the variation of the energy with strain (see Fig. \ref{fig:1} for an illustration). These characterisations can be made in principle in material specific ways. 

For the present work, I choose the stability range to be $\gamma_{\pm} = \gamma_0 \pm \sqrt{-E_0}$, and the energy at a given strain within the stability range as $E(\gamma,E_0,\gamma_0) = E_0 + {\mu \over 2} (\gamma - \gamma_0)^2 $. These choices embody the expectation  (supported by numerical evidence \cite{Bhaumik2019}) that the stability range as well as the elastic energy increase before instability increase upon lowering the minimum energy. I consider a Gaussian distribution of energies, $P_{0}(E_0) = \sqrt{2 \over \pi \sigma^2} \exp(-{E_0^2 \over 2 \sigma^2})$, for $-1 < E_0 < 0$, $\sigma = 0.1$ \footnote{This choice, for concreteness, would correspond to the density of states of a subvolume of $50$ particles, with the number of states $\Omega \sim \exp(\alpha N)$, $\alpha \sim 1$ \cite{Sastry2001}}.

For a given energy $E_0$, it is assumed that one has mesostates with several possible values of the  {\it stress-free} strain $\gamma_0$, reflecting the fact that distinct configurations of the same energy may exist, related to each other by a shift in the strain values at which they are stable. \footnote{This in principle introduces an additional $E_0$ dependent in the density of states, but we consider an implementation where the total weight of all relevant mesostates of a given energy is given by $P_{0}(E_0)$.}. For concreteness, two specific cases are considered here: (1) The {\it regular} case, where mesostates have $\gamma_0 = n \times 2 \sqrt{-E_0}$, $n$ being an integer, so that mesostates at a given energy are present in a periodic, non-overlapping fashion.  (2) The {\it uniform} case, where the $\gamma_0$ values are uniformly distributed.

\begin{figure}[htp]
\centering{}
\includegraphics[scale=0.24]{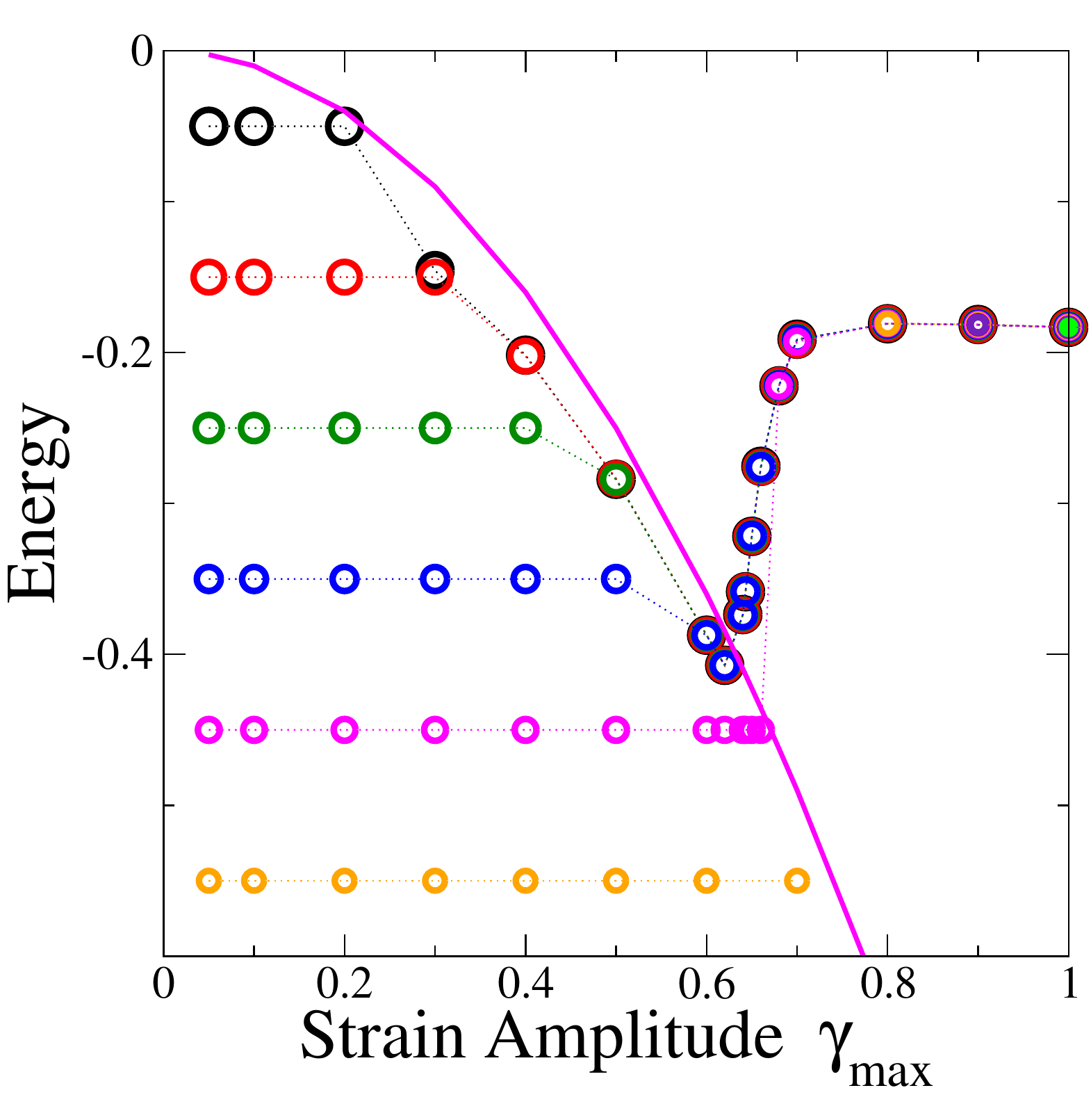} 
\includegraphics[scale=0.24]{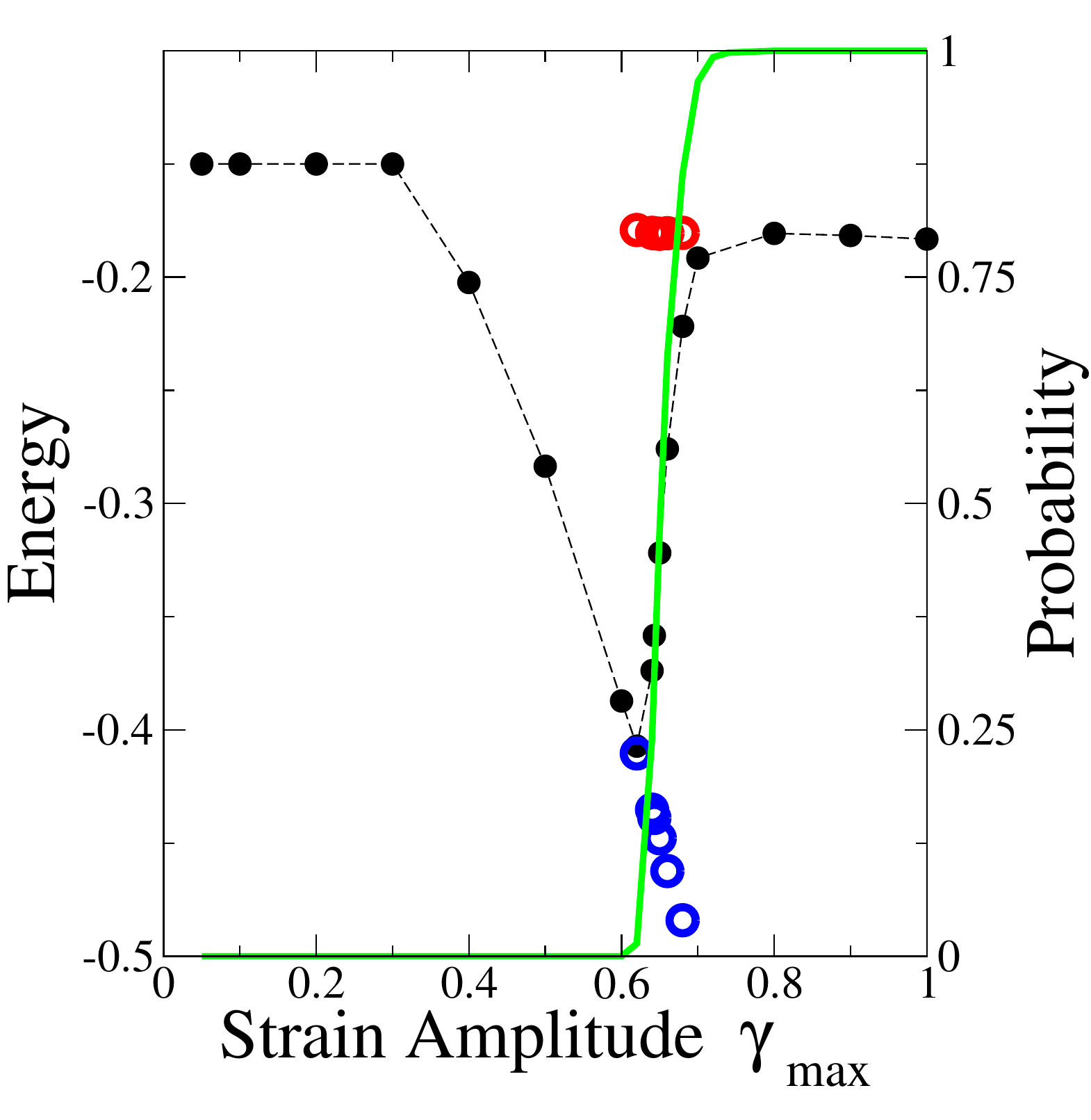} 
\caption{(Left) The yielding diagram that shows the steady state energies reached after repeated cyclic deformation, as a function of the strain amplitude $\gamma_{max}$, for a range of initial energies. (Right) Along with average energies over all samples (black), partial average energies are shown, of samples that yield (red) and those that do not (blue), revealing the discontinuous nature of the transition to the yielded state. The yielding probability {\it vs.} strain is also shown  (green).} 
\label{fig:2}
\end{figure}

When the stability limit $\gamma_{\pm}$ is reached, the state of the system makes a transition from the current mesostate to another. In the athermal case, clearly, such a transition is possible only to other states with lower energy {\it at the same strain value} (see Fig. \ref{fig:1}). Depending on the $E_0$ and $\gamma_{0}$ values of the other states, a transition may occur to mesostates with higher or lower $E_0$, a key feature on which the behaviour described below depends (which requires $\mu > 1$ for the {\it regular} case).  One may consider a transition rule that permits  transitions to (i) {\it any} mesostate with a lower energy at the transition strain, or as a physically motivated choice, (ii) restrict the range of $E_0$ values to which transitions are permitted, within a range $\delta E$. Both these cases are considered. 

\begin{figure}[htp]
\centering{}
\includegraphics[scale=0.24]{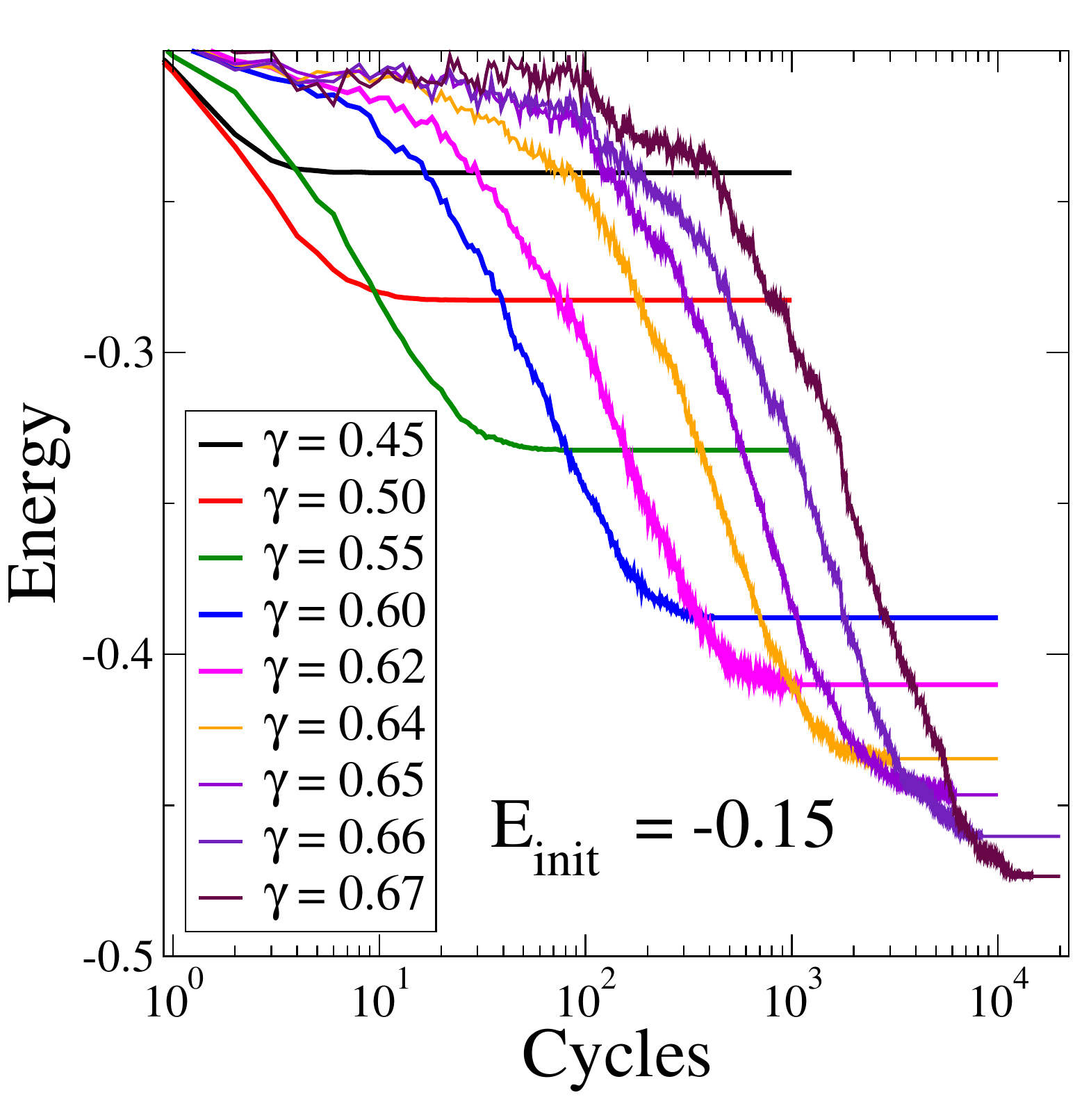} 
\includegraphics[scale=0.24]{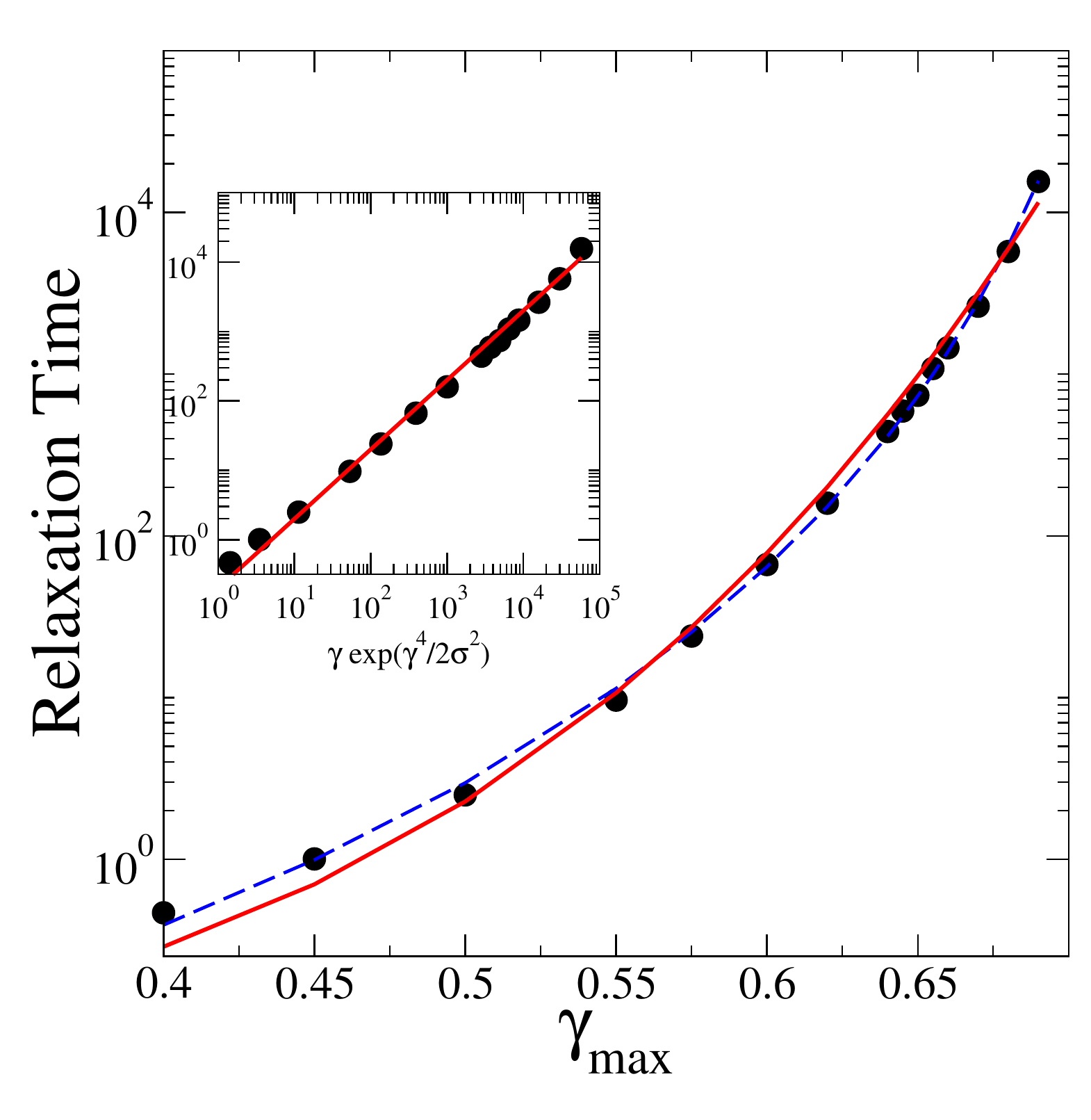} 
\caption{(Left Panel) Average energies as a function of the number of cycles. (Right Panel) Relaxation time $\tau$ to reach the steady state below yielding. The dashed line is a power law fit, and the solid line describes $\tau = A~  \gamma_{max} \exp({\gamma_{max}^4 / 2 \sigma^2})$. The inset shows a log-log plot of $\tau$ {\it vs.} $\gamma_{max} \exp({\gamma_{max}^4 / 2 \sigma^2})$. } 
\label{fig:3}
\end{figure}

The case of {\it regular} distribution of $\gamma_0$ is considered first, with a choice of $\mu = 1.1$ and investigated numerically as follows. Starting with an initial energy $E_0$, and $\gamma_0 = 0$, the strain $\gamma$ is varied cyclically with amplitude $\gamma_{max}$, and the state of the system is propagated untill a stability limit $\gamma_{\pm}$ is reached. When this happens, the set of all states to which a transition can occur, with the corresponding weights (given by $P_{0}(E_0)$; in practice, the $E_0$ range is divided into $10^4$ bins) is evaluated, and a new state, with its $E_0$ and $\gamma_0$ values, is randomly chosen among permitted states. The procedure is repeated for $10^3$ cycles of shear. If the state of the system does not change at the end of successive cycles, the procedure is terminated. Otherwise, the zero strain energies for the last $200$ cycles are averaged. This procedure is repeated for $10^3$ independent sample runs.  Fig. \ref{fig:2} (a) shows the resulting final energies as a function of both a set of initial energies and strain amplitudes. Strikingly, all the essential features of observations from simulations are reproduced. For high initial energy values, energies shift to lower energies, towards a threshold value common to a range of initial energies, before a discontinuous transition occurs, after which the energies follow a common branch regardless of initial energy values. For initial energies lower than the threshold, there is no change {\it vs.} $\gamma_{max}$ till a yield value is reached, at which a discontinuous jump in energy is seen. Naturally, the discontinuity in energy grows with the reduction of the initial energy. From  Fig. \ref{fig:2} (a), the discontinuous nature of the transition is not fully apparent.  To reveal it more clearly, for any choice of $\gamma_{max}$, the final energies for the cases where a transition occurs (and those for which it does not) are computed, as also the probability of occurrence of the transition {\it vs.}  $\gamma_{max}$. These are shown for $E_{0}^{init} = -0.15$ in  Fig. \ref{fig:2} (b), which illustrates that (i) the yielding probability rises sharply from $0$ to $1$ in a very narrow range of $\gamma_{max}$, and (ii) the energy averaged over cases where a transition does not occur continues to decrease with $\gamma_{max}$ even beyond the yield strain range, whereas when a transition occurs, a nearly constant final energy is reached. The latter observation, interestingly, echoes simulation results \cite{parmar2019strain} where continued annealing is observed beyond the yield strain amplitude, away from the shear band in which strain gets localised. To interpret the results here suitably, one must consider that the present system represents the behaviour of one mesoscale block in a more extended system. Better still, such an extended system should be investigated, which will be taken up in future work. A comment on the average energies beyond yielding is also in order in this context. From the results here, the post-yield energy does not continue to increase, whereas in simulations, it does. This can be simply interpreted as a result of a larger fraction of the simulated system undergoing yielding rather than a change in the typical energy within that subvolume. Results in \cite{parmar2019strain} clearly support such an interpretation.

Next, the energies of individual samples {\it vs.} cycle number averaged for $\gamma_{max}$ below the yield value,  are shown in Fig. \ref{fig:3} for an initial energy of $-0.15$. The time scales (measured in terms of cycles) to reach the final state increase with $\gamma_{max}$ consistently with simulation results \cite{Fiocco2013,leishangthem2017yielding}, and can be fitted to a power law as shown, but the fit parameters are not meaningful. A way of understanding the increasing time scales is presented below, which leads to the expectation that $\tau \sim  \gamma_{max} \exp({\gamma_{max}^4 \over 2 \sigma^2})$ which provides a very reasonable description. No long relaxation times are observed above the transition, which I interpret to indicating that the long relaxation times observed in simulations arise from parts out the system outside the yielded region rather than the yield region itself, which can be tested in simulations. 

\begin{figure}[htp]
\centering{}
\includegraphics[scale=0.24]{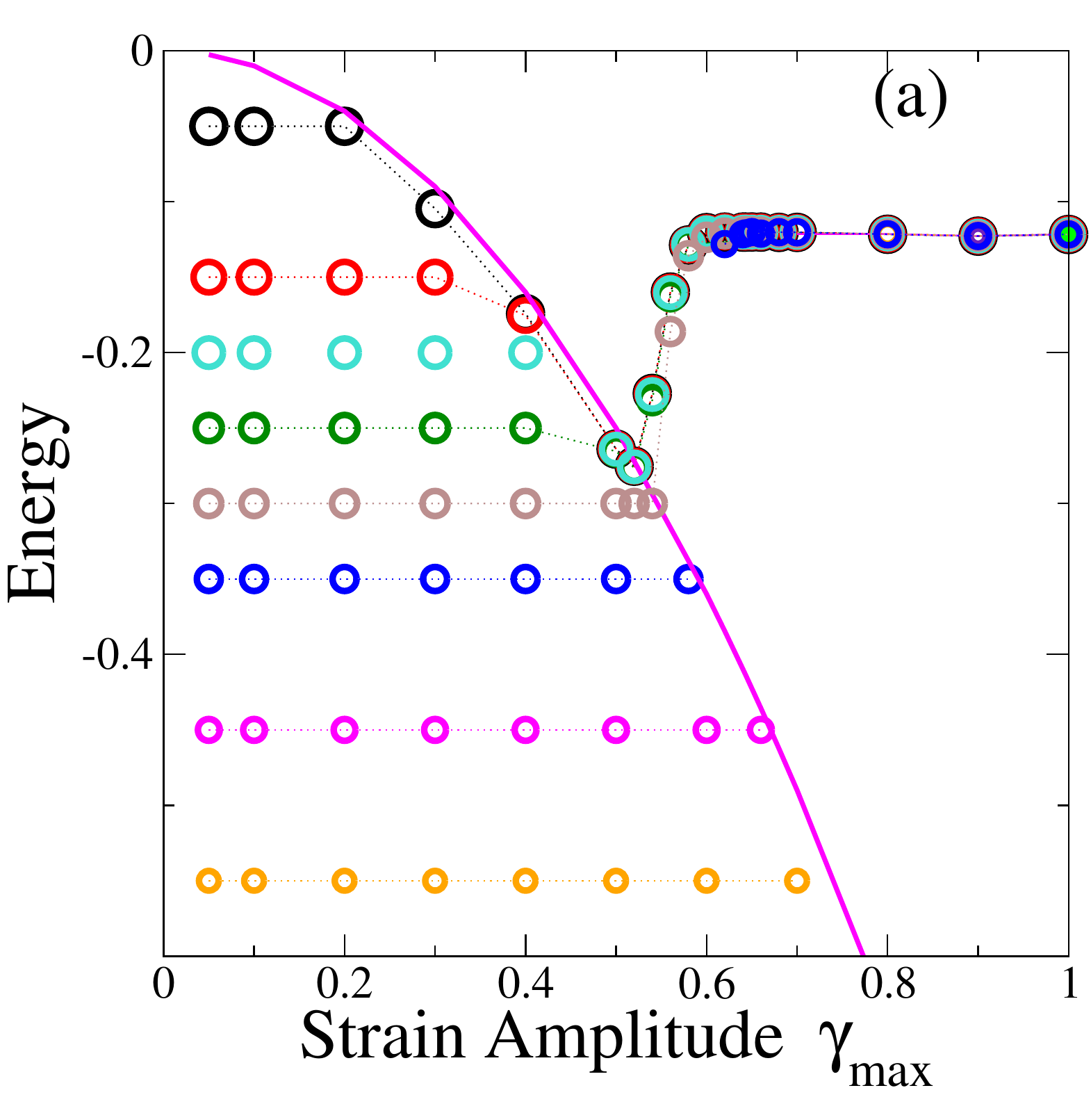} 
\includegraphics[scale=0.24]{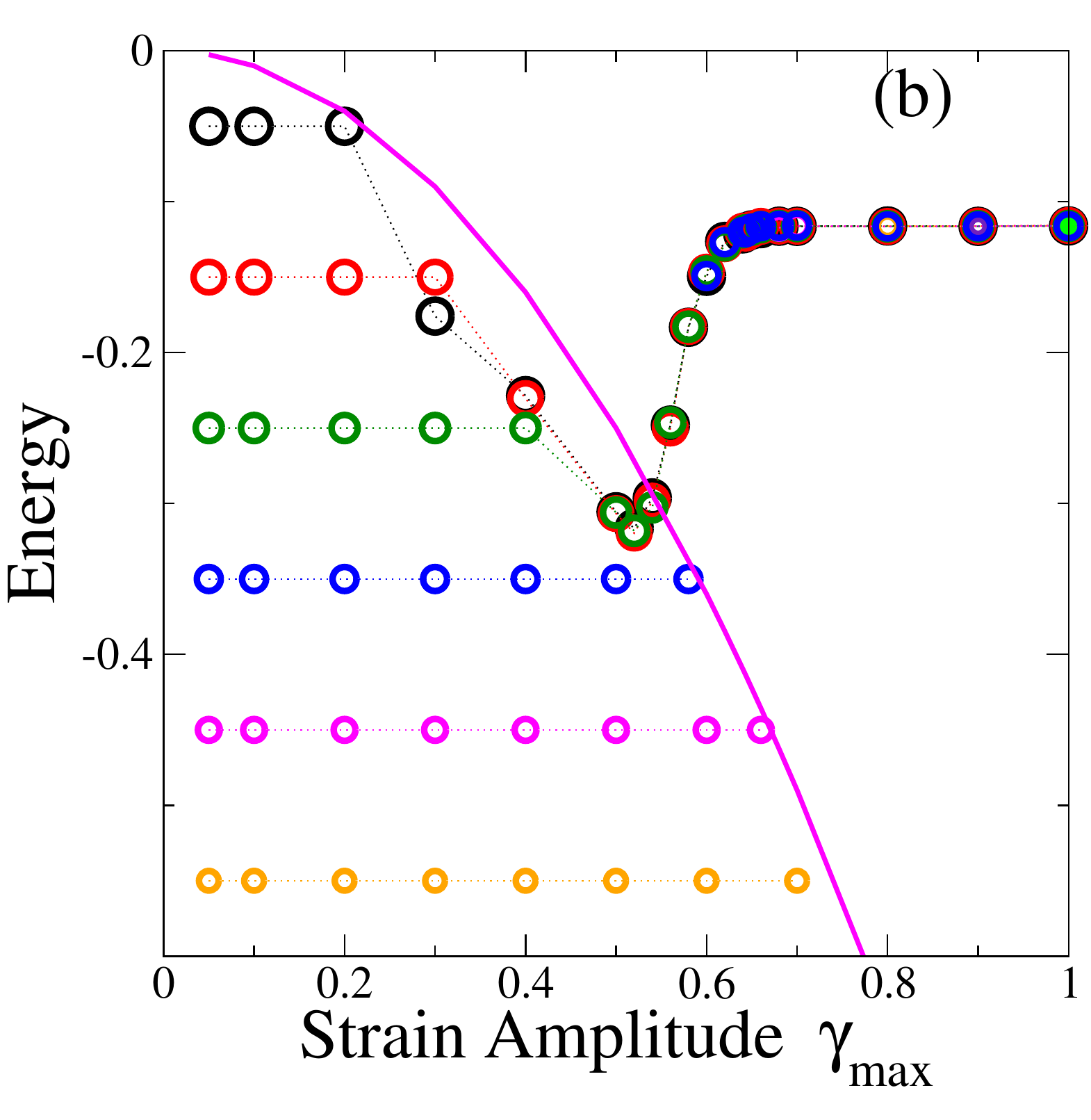} 
\includegraphics[scale=0.24]{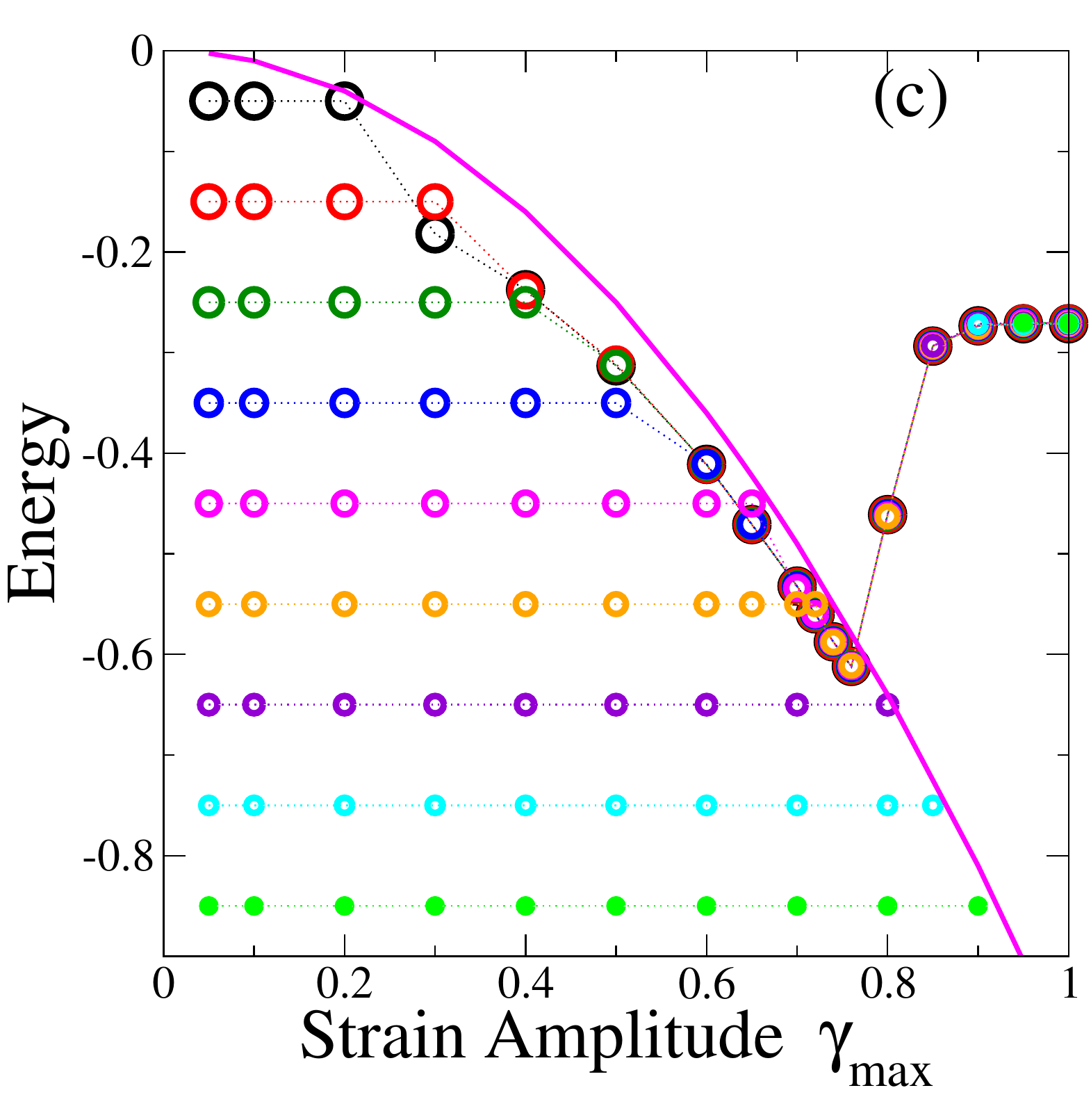} 
\includegraphics[scale=0.24]{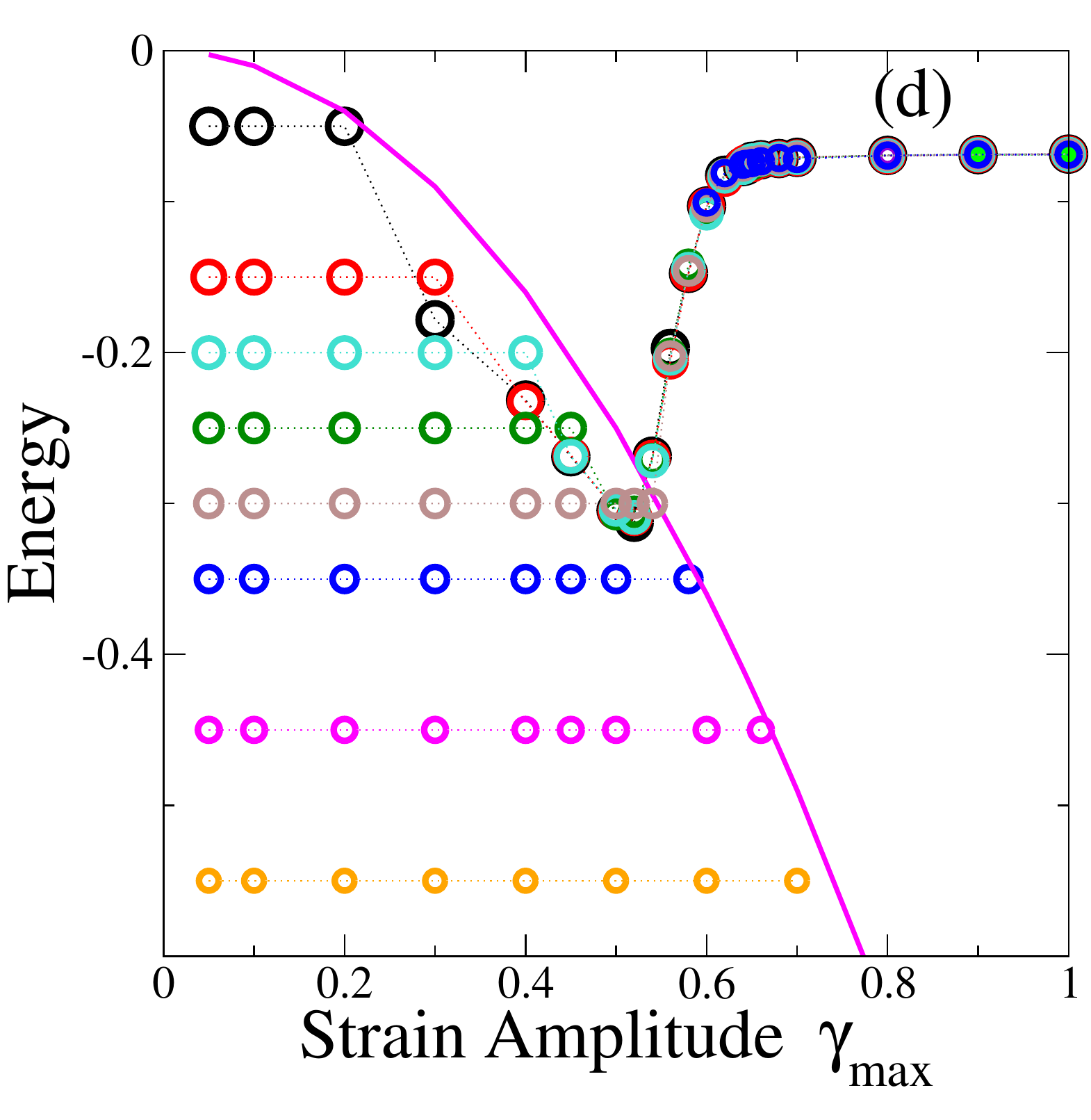} 
\caption{Yielding diagram for different cases: (a) The {\it constrained} case with $\delta E = 0.05$,  $\sigma = 0.1$,  $\mu = 1.1$ . (b) The {\it uniform} case with $\sigma = 0.1$,  $\mu = 1.1$  (c) The {\it regular} case with $\sigma = 0.15$,  $\mu = 1.1$ (d) The  {\it SGR} version, with $\sigma = 0.1$, $\mu= 2$.} 
\label{fig:4}
\end{figure}

Before discussing the results further, I consider variations of the model described above. The first is the {\it regular} model discussed above, but with a constraint on the range of energy change when a mesostate transition occurs (with the choice $\delta E = 0.05$,  $\sigma = 0.1$,  $\mu = 1.1$). The second is the {\it uniform} model ($\sigma = 0.1$, $\mu = 1.1$) with $\gamma_{0}$ values uniformly distributed, rather than discretely, for any $E_0$. I also consider the {\it regular} case with $\sigma = 0.15$,  $\mu = 1.1$.  Finally, with  $\mu = 2$, which permits transitions to all mesostates without regard to their energy (since the threshold energy for all states is $0$), and with the additional condition that the $\gamma_{0}$ value for the new state is the same as the current value of strain, one realises a specific instance that is the same as the SGR model \cite{sollich1997rheology}. The relevant yielding diagrams are reported Fig. \ref{fig:4}. Investigation of these variations reveals that qualitatively, each of them reproduces the behaviour discussed above. Equally importantly, these variations do change the threshold energy as well as the energy beyond yielding, which may be relevant in understanding material specific behaviour. 
 The {\it constrained} case demonstrates that transitions to widely different states is not a necessary feature to produce the observed behaviour. The {\it regular} case with $\sigma = 0.15$ illustrates that the distribution $P_{0}(E_0)$ has a strong bearing on the threshold energy. Importantly, it also suggests that models with a trivial ({\it e. g.}, flat) distributions in the stability range are likely not to capture the yielding transition under athermal cyclic shear, which should be tested. Indeed, for $\sigma = 0.25$, the simulations converge to energies below $-\gamma_{max}^2$ all the way to the lowest energies considered. The results indicate that in the present formulation, limit cycles associated with memory effects \cite{Adhikari2018,Mungan2019} do not arise.  Constructing suitable transition graphs \cite{Mungan2019,MunganWitten2019,Fiocco2015} that can describe limit cycles associated with memory effects  need to be explored.

\begin{figure}[htp]
\centering{}
\includegraphics[scale=0.33]{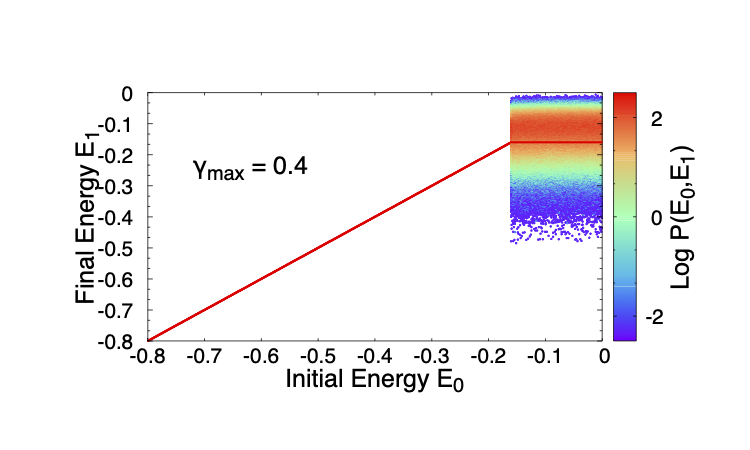}
\includegraphics[scale=0.33]{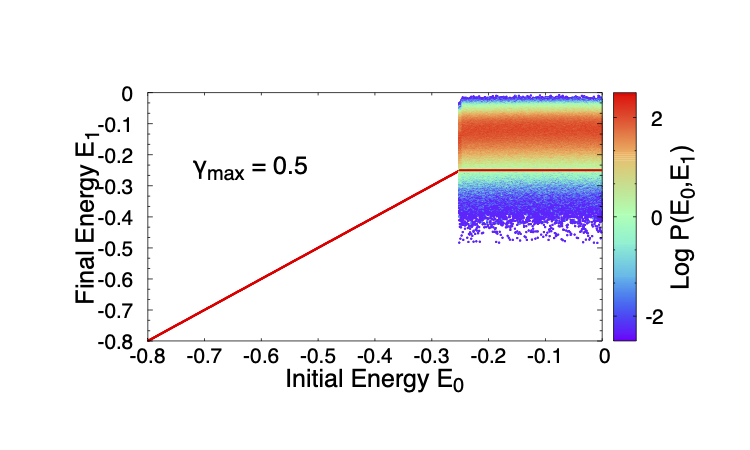}
\includegraphics[scale=0.33]{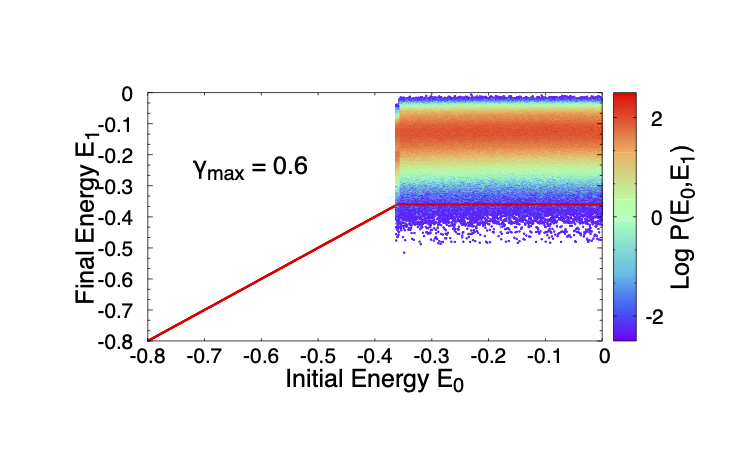}
\includegraphics[scale=0.33]{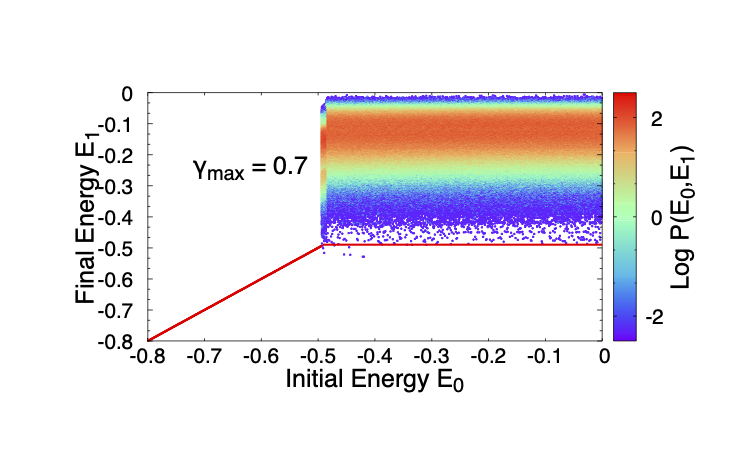}
\caption{The histogram of energies reached after one cycle (y-axis) {\it vs.} the initial energy (x-axis) for a range of strain amplitudes $\gamma_{max}$ indicated in the panels, for the {\it uniform} case with $\mu = 1.1$. The yield $\gamma_{max}$ in this case is $\approx 0.563$.} 
\label{fig:5}
\end{figure}

The most intriguing result here and in earlier simulations is the presence of the threshold energy, which is not an obvious part of the model description. In contrast, the mean energy beyond yielding has the simple explanation. It is the energy above which the mean energy {\it to which} a mesostate transitions is lower, whereas below, it is higher.  To investigate the meaning of the threshold energy, I consider, given a strain amplitude $\gamma_{max}$, and initial energy, the distribution of energies to which the system transitions, {\it at the end of a full cycle}, for the {\it uniform} case with  $\mu = 1.1$. Such two dimensional distributions $P(E_0,E_1)$ are shown for a series of $\gamma_{max}$ values in Fig. \ref{fig:5}.  For any $\gamma_{max}$, there is trivially a range of energies for which no transition takes place, but for energies above,  the transition probabilities are roughly independent of the initial energy (though this feature is not common to all the cases studied, behaviour in other cases is qualitatively the same). It can be seen that the probability to transition to an energy below the threshold for the given strain ($ = -\gamma_{max}^2$, indicated by a horizontal line) decreases dramatically with an increase in $\gamma_{max}$, and, between $\gamma_{max} = 0.5$ and $0.7$, it drops strongly (the strain at which the yielding probability is $1/2$ in this case is $\approx 0.563$). Thus, the probability to transition to the yielded state becomes overwhelming, but is never equal to $1$. Correspondingly, the probability over a large number of cycles to transition to the yielded state becomes a sharper function, but moves to higher strain values. 

Keeping in mind the case when $\mu = 2$ wherein an unstable mesostate can make a transition to any other mesostate without restriction, a simple, if approximate estimate of the transition probability to yielding can be made as follows. Since the stability range for a state of energy $-E_0$ is $2 \sqrt{-E_0}$, over a cycle of strain (or any unit of strain), the transition rate out of that state is inversely proportional to $\sqrt{-E_0}$, and the probability to transition to a new state of energy $E^{'}$ is $P_{0}(E^{'})$. If one assumes that as a result of these transitions, a stationary distribution of reached within a single cycle, the stationary distribution is given by  $P_{st}(E_0) \propto \sqrt{-E_0} P(E_0)$. This is inaccurate for low energies, and also does not incorporate the fact that once an energy below $-\gamma_{max}^2$ is reached, no further transitions occur. Nevertheless, it is a good approximate description of the distribution of energies reached after one cycle.  One can evaluate the probability, after one cycle, of getting trapped in a minimum that is stable at $\gamma_{max}$, $p_{trap} (\gamma_{max}) = \int_{-\infty}^{-\gamma_{max}^2} P_{st}(E_0) dE_0 = \Gamma({3\over4},{\gamma_{max}^4 \over 2 \sigma^2})/\Gamma({3\over4})
\sim \exp(-{\gamma_{max}^4 \over 2 \sigma^2})/ \gamma_{max}$.  The relaxation time should go inversely as $p_{trap}$,  $\tau \sim  \gamma_{max} \exp({\gamma_{max}^4 \over 2 \sigma^2})$,
which was seen to be  good description of the simulated results (Fig. \ref{fig:3} (b)). Writing the probability after $N_{cyc}$ cycles to yield as $p_{trans} = (1 - p_{trap})^{N_{cyc}} \sim \exp(-p_{trap} N_{cyc})$ we see that the 
$p_{trans}$ will be an increasingly sharp function of $\gamma_{max}$ and the yield strain amplitude, $\gamma_{max}^{Y} \sim (\log(N_{cyc}))^{1/4}$, depends very weakly on $N_{cyc}$. Whether interactions between mesoscale regions in a macroscopic sample could lead to a sharp transition needs to be investigated. In \cite{Bhaumik2019}, it was observed that the threshold energy closely corresponds to the energy at which the average index of saddle points goes to zero, marking a change in the character of the energy landscape. The implications of such a qualitative change, and how it may be incorporated in the present description, merits further investigation. 

In summary, the yielding behaviour under athermal cyclic deformation has been investigated for a family of models envisaged to represent a mesoscopic system, which remarkably captures several key observations for model glasses that have been investigated through computer simulations using athermal quasistatic shear. The mechanism for the transition is a competition of the stability against cyclic shear of low energy mesostates, and the {\it entropic drive} associated with the large number of higher energy minima present, reminiscent of arguments presented in \cite{Itamar2016yielding,parisi2017shear}. The transition is always discontinuous, as also noted for uniform shear in a recent theoretical analysis \cite{Fielding2020}.These suggest that such a description, embedded in an elasto-plastic scheme is suitable for further investigation as an approach to modelling the larger length scale behaviour of sheared amorphous solids, such as the emergence of permanent or multiple shear bands \cite{vasisht2018permanent}. As the approach is based on a local energy landscape description, it has advantages in capturing thermal behaviour as well as athermal behaviour. It also suggests detailed investigation of the energetic aspects of local plastic events in simulations \cite{pinaki2020} in order to understand better how a diversity of amorphous solids may be satisfactorily modelled. 

\begin{acknowledgments}
I thank Vishwas Vasisht, Muhittin Mungan, Pallabi Das, Monoj Adhikari and Surajit Sengupta for comments on the manuscript, Yagyik Goswami with help with manuscript preparation, and support through the J. C. Bose Fellowship, SERB, DST, India.
\end{acknowledgments}








\bibliographystyle{apsrev4-1}
\bibliography{mesobib}

\end{document}